\documentstyle[12pt,psfig]{article}
\begin{document}
\def\e{\mbox{e}}
\def\sgn{{\rm sgn}}
\def\gsim{\;
\raise0.3ex\hbox{$>$\kern-0.75em\raise-1.1ex\hbox{$\sim$}}\;
}
\def\lsim{\;
\raise0.3ex\hbox{$<$\kern-0.75em\raise-1.1ex\hbox{$\sim$}}\;
}
\def\MeV{\rm MeV}
\def\eV{\rm eV}
\thispagestyle{empty}

\vskip 0.3cm

\LARGE
\begin{center}

{\bf Can we distinguish Majorana and Dirac neutrinos in solar neutrino
experiments ?}
\end{center}
\normalsize
\vskip1cm
\begin{center}
{\bf V.B. Semikoz}
\footnote{E-mail: semikoz@charley.izmiran.rssi.ru"~, fax: 7-095-3340124}
\end{center}
\begin{center}

  {\small {\em The Institute of the Terrestrial Magnetism, the Ionosphere and
Radio Wave Propagation of the Russian Academy of Sciences,}}

  {\small {\em IZMIRAN,Troitsk, Moscow region, 142092, Russia}}
\end{center}

\begin{abstract}

If neutrino conversions within the Sun result in partial polarization
of initial solar neutrino fluxes then a new opportunity to
distinguish Majorana and Dirac neutrinos by measuring the differential
$\nu_ee$-scattering cross section in solar neutrino detectors arises.

The experiment like HELLAZ would be preferable in testing
of recoil electron spectra differences initiated by
Majorana and Dirac neutrinos since low energy recoil electrons
($T\geq 100~keV$) can be detected and electron energy and direction
can be determined with good precision.



\end{abstract}

\newpage

\section{Introduction}
Lower bounds on neutrino masses have not been found yet in direct laboratory
experiments. In particular, in the case of neutrinoless double-beta
decay one expects a decrease of an upper bound on
the Majorana mass $m_{\nu_e}^{(M)}$ only.  This current improvement of
upper limits on neutrino masses takes a long time and strong efforts.
 However, we can not justify on this way whether neutrino is really the
 Majorana particle or it can be the Dirac one, or a mixture of them
 (ZKM-neutrino).

In this connection let us recall the old experiment by Davis\cite{Davis} who
demonstrated that neutrino and antineutrino are indeed different particles
if we are using $\tilde{\nu}_e$
from the beta-decay $n\to pe^-\tilde{\nu}_e$ as the incident "neutrino"
for the capture process $\nu_e + \mbox{}^{37}Cl\to \mbox{}^{37}Ar +
e^-$.

Negative result of the experiment\cite{Davis}, $\nu_e\neq
\tilde{\nu}_e$, is not an evidence that $\nu_e$ and $\tilde{\nu}_e$ are
 the Dirac neutrinos with the fixed lepton number $L = \mp 1$.
In such experiments the helicity $r=\mp 1$ (upper signs for $\nu_e$)
is appropriate quantum number which is conserved
due to the $V-A$ law of charged current weak interaction rather the
lepton number.

Both the right-handed Majorana neutrino and the Dirac antineutrino with
the same helicity $r= +1$ could be emitted in the beta decay
$n\to pe^-\tilde{\nu}_e$
with the following suppression of the spin-flip $r = +1\to r = -1$ in the
process
of capture in the chlorine detector (the latter is true to the order
of $O((m_{\nu}/E)^2)\ll 1$ in the cross section).

Thus this example demonstrates the well-known fact that {\it in the massless
limit $m_{\nu}\to 0$
Majorana and Dirac neutrinos are not distinguishable}. We can not mark
any difference between the fully-polarized right-handed Majorana
neutrino $\tilde{\nu}_{e_R}$ and the right -handed Dirac  antineutrino
 $\tilde{\nu}^{(D)}_e$ as well as between the left-handed Majorana
neutrino $\nu_{e_L}$ and the Dirac left-handed one, $\nu_{e_L} =
 \nu_{e_L}^{(D)}$ (see below section 3).

In turn, if an incident neutrino flux became
{\it partially-polarized} this  would give a chance to
distinguish these particles.

We propose here a new way for distinction of Majorana and
Dirac neutrino in the solar neutrino experiments by studying the profiles of
the electron spectra
in the low-energy $\nu_ee$-scattering for incident $\nu^{(M)}$ and
$\nu^{(D)}$ fluxes.  It seems possible when solar neutrino flux is
partially-polarized.

The ultrarelativistic neutrinos produced in thermonuclear reactions
within solar interior are evidently the left-handed ones (fully-polarized
$\nu_{e_L}$) and one needs to assume some mechanism for their conversion to
 the right-handed neutrinos.

First, the conversions $\nu_{e_L}\to \tilde{\nu}_{e_R}$ in
  the Majorana case or $\nu_{e_L}\to \nu_{e_R}$  in the Dirac case are
obviously based on the assumption of a non-vanishing neutrino mass
$m_{\nu}\neq
0$. This leads to nonzero neutrino diagonal \cite{Shrock1} and
transition \cite{SchechterValle} dipole moments and, therefore, to the
possible neutrino spin \cite{Shrock1} and spin-flavor
precession \cite{VVO} in vacuum in the presence of an external magnetic field.
In a medium neutrino oscillations and spin-flavor precession can occur
as the resonant conversions $\nu_{e_L}\to \nu_{\mu_L}$
\cite{MikheevSmirnov} and $\nu_{e_L}\to \tilde{\nu}_{\mu_R}$
\cite{Akhmedov}.

The spin-flavor conversion in
combination with the MSW-mechanism can lead to the right-handed
Majorana neutrino production ($\nu_{e_L}\to \tilde{\nu}_{e_R}$
\cite{APS}, see below section 2), i.e. to a mixture of the left-and
right-handed active neutrinos as a partially-polarized
$\nu_{e_L}~,\tilde{\nu}_{e_R}$ neutrino flux incident upon underground
detectors. In contrast to the Majorana neutrino case, for the same
conditions in the Sun the right-handed Dirac neutrinos produced via the
spin-flip $\nu_{e_L}\to \nu_{e_R}$ or in the cascade conversions
$\nu_{e_L}\to \nu_{\mu_R}\to \nu_{e_R}$ appear to be sterile ones
with respect to the $\nu_ee$-scattering in detectors.

Notice that
necessary large values of transition moments (even
without direct proportionality to $m_{\nu}$) can be obtained in
some extended models obeying all known laboratory, astrophysical and
cosmological constraints on neutrino masses and on its dipole moments.

For all Majorana and Dirac neutrinos with a mass below $5~keV$ the
most restrictive limit on dipole or transition magnetic and electric
moments $\mu_{\nu}\lsim 3\times
10^{-12}\mu_B$ arises from the absence of anomalous neutrino emission
from the red-giant cores just before helium ignition\cite{Raffelt}.
The condition $m_{\nu}\lsim 5~keV$ follows from a kinematic limit on
the neutrino energy $E_{\nu_{a,b}}$ lost in the plasmon decay
$\gamma^*\to \nu_a\tilde{\nu}_b$ since the plasma frequency in a
degenerate electron gas of red- giants is bounded in the same region,
$\omega_p\lsim 5-10~keV$.

A large Dirac neutrino magnetic moment (both
diagonal and transition ones, $\mu_{\nu}^{(D)}\gsim 3\times
10^{-12}\mu_B$) was also excluded from SN1987A neutrino events in the\\
Kamiokande and IMB detectors. This is due to non-observation there of
a hard energy tail stipulated by the sterile $\nu_{e_R}$ emission from
a hot supernova core\cite{Barbieri1}. These neutrinos could be produced
within core via the electromagnetic scattering (see below Eq.~(\ref{em}))
and then be converted to the active $\nu_{e_L}$ in the intergalactic
magnetic field .

The absence of
SN1987A constraint in the Majorana case means that the model-dependent
estimate of $\mu_{e\mu}$ \cite{Raffelt} seems to be less consistent
even for light neutrinos suitable for the resonant spin-flavor or
the MSW conversions in the Sun.

Therefore the laboratory constraint from reactor antineutrino
experiments which is common for diagonal and transition magnetic
moments, $\mu_{\nu_e}\lsim 2-4\times 10^{-10}\mu_B$ \cite{Kuld},
remains an upper estimate of Majorana neutrino transition moments
corresponding to effective neutrino conversions within solar convective
zone with magnetic fields of order $\sim 1~kG$.

On the other hand, in magnetic hydrodynamics one can not exclude
solutions with a strong magnetic field near bottom of the
convective zone of the Sun,  $\sim 100~kG$\cite{Hughes}, and even
larger within solar core for equilibrium hydromagnetic configuration in
the gravitational field, $B\lsim 10^8~G$\cite{Solov'ev} .  As a result
even for the case when the limit $\mu_{e\mu}\lsim 3\times
10^{-12}~G$\cite{Raffelt} is valid we may apply some mechanisms for
effective spin-flavor conversions.

Notice also that the most stringent constraints on transition magnetic
moments\\
$\mu_{es}\lsim 10^{-16}\mu_B$ were derived in \cite{Pastor} from
the primordial nucleosynthesis bound on additional neutrino species and
from a supernova energy loss argument. These bounds were obtained
in \cite{Pastor} for {\it active-sterile neutrino conversions neglecting
neutrino mixing} and in a medium with random magnetic fields.
Therefore this approach has no relation to our consideration here
since we are using the models \cite{Akhmedov,APS} {\it for
active-active neutrino conversions}.

In the end of section 2 we consider
experimental bounds on $\tilde{\nu}_{e_R}$ and give a theoretical
 interpretation of some contradictions in the model used.
 In particular, we consider an important question how
to avoid some known lacks of the spin-flavor scenario
\cite{APS} with the use of large neutrino magnetic moments that inevitably
leads to contradiction with non-observation of temporal variations of
the solar neutrino flux in the most of current experiments.

Then in section 3 we show why
the scattering of partially-polarized neutrino flux off electrons
should be absolutely different for Majorana and Dirac neutrino provided
that their right-handed components interact (do not interact in the
Dirac case) with electrons.

In section 4 we discuss results and crucial parameters of the model
used.

\section {Mechanisms of the right-handed Majorana and
Dirac neutrinos production in the Sun.}

The main assumption here is the presence of the right-handed neutrinos
$\tilde{\nu}_{e_R}$, $\nu_{e_R}$  produced by some mechanism within the
Sun.  For instance, in the spin-flavor scenario involving Majorana
neutrinos the Sun would be a source for antineutrinos, some of which
could be $\tilde{\nu}_{e_R}$'s by a combination of spin-flavor
($\nu_{e_L}\to \tilde{\nu}_{\mu_R}$) and flavor ($\tilde\nu_{\mu_R}\to
\tilde{\nu}_{e_R}$) oscillations (or due to permutation of steps above,
$\nu_{e_L}\to \nu_{\mu_L}\to \tilde{\nu}_{e_R}$ ). This scenario was
in detail elaborated  in \cite{APS} for the case of twisting magnetic
field that allowed authors to avoid the suppression of the cascade
conversion $\nu_{e_L}\to \tilde{\nu}_{e_R}$ caused by the almost full
compensation of the partial (two-step) amplitudes, $M_1 + M_2\approx
0$, (for a small neutron abundance in the Sun).

Moreover, authors
\cite{APS} found conditions when either $\nu_{e_L}$-
$\tilde{\nu}_{e_R}$-system decouples from $\nu_{\mu_L}$ and
$\tilde{\nu}_{\mu_R}$ or the triple resonances merge and a resonant
transition leads to a complete
$\nu_{e_L}\to \tilde{\nu}_{e_R}$-conversion. In the case of separation of
$\nu_{e_L}$- $\tilde{\nu}_{e_R}$-system the adiabaticity condition is
not fulfilled (at least for the convective zone in the Sun) while for
the triple resonance  the relevant adiabaticity parameters at
the merging point depend on neutrino energy.

As result authors\cite{APS} predict that the $\tilde{\nu}_{e_R}$-flux
should have a peak at the energy $E$ defined by the resonant condition.
Explicitely this occurs for $\dot{\Phi}< 0$ at $(s_2\delta)
\simeq \mu B_{\perp}$, where $s_2 = \sin 2\theta$ is the mixing
parameter; $\delta = \Delta m^2/4E$ and  $\Delta m^2>0$ is the
neutrino squared mass difference; $\mu$ is the neutrino transition
magnetic moment; $B_{\perp}$ is the amplitude of the twisting magnetic
field {\bf B}$\sim e^{i\Phi(t)}$ which is perpendicular to the
momentum {\bf k} of an ultrarelativistic neutrino, $E\approx k$.

Since the position and the width of the peak strongly depend
 on the magnetic field strength in the region of resonance, the energy
spectrum of resonantly emitted solar $\tilde{\nu}_{e_R}$'s should
exhibit characteristic time dependence. In particular, depending on
the signs of $\dot{\Phi}$ in the northern and southern
solar hemispheres the semiannual neutrino flux variation\footnote{First
these semiannual variations were predicted in \cite{VVO} for the
neutrino magnetic moment interaction with the constant toroidal
magnetic field (azimuthal component) which changes sign at the solar
equator.  Inclination of the ecliptic (of the Earth orbital plane) with
respect to solar equator ($7^{\circ}~15'$) leads to the minimum of
the neutrino interaction term $\mu B_{\perp}$ for the
two season positions of the Earth (in June and December) when the solar
core is viewed from the Earth through the solar equator.}
can be either enhanced or suppressed.  Moreover,
the energy at which the $\tilde{\nu}_{e_R}$-spectrum achieves the
maximum should vary in time .

Note also other ways of the Majorana $\tilde{\nu}_e$-production like
$\nu_e\to \tilde{\nu}_e + \chi$-decay (to majoron) enhanced in solar
matter \cite{Berezhiani}. In this
case we do not need any strong magnetic field in the Sun or the
presence of a large neutrino magnetic moment. However, the
$\tilde{\nu}_e$-spectrum differs in such case  from the initial one
for left-handed neutrinos that prevents from the approach developed
below.

The active Dirac antineutrino $\tilde{\nu}_{e_R}= \tilde{\nu}_e^{(D)}$
never can  be produced in the Sun due to absence there of the
annihilation process $e^-e^+\to \nu \tilde{\nu}$ which is important,
for instance, in a supernova. However, in the same twisting magnetic
field the resonant $\nu_{e_L}^{(D)}\to \nu_{e_R}^{(D)}$-conversions
lead to the production of the sterile neutrinos $\nu_{e_R}^{(D)}$
through a diagonal magnetic moment $\mu_{\nu}$\cite{Smirnov} (
neglecting neutrino flavor mixing) or  analogously to the Majorana case
in the presence of mixing and via a transition magnetic moment for the
system of the active $\nu^{(D)}_{e_L}$, $\nu^{(D)}_{\mu_L}$ neutrinos
and their sterile $\nu^{(D)}_{e_R}$, $\nu^{(D)}_{\mu_R}$- components.

On the other hand, it is well-known that the Kamiokande
data already yield restrictive limits on a solar
$\tilde{\nu}_{e_R}$-flux at the high energy region, $E_{\nu}\gsim
8.5~MeV$ \cite{Suzuki}. This limit was found from an isotropic background
accounting
also for $\tilde{\nu}_{e_R}$'s due to the relatively large
cross-section $\sigma (\tilde{\nu}_ep\to ne^+)\sim 9.4\times
10^{-44}~cm^2(p_eE_e/MeV^2)$ with the isotropic distribution of
positrons, where $E_e = E_{\nu} - 1.3~MeV$.  The best relative limit is
at $E_{\nu} = 13~MeV$ where the $\tilde{\nu}_{e_R}$ flux is less than
5.8 \% of $^8B$ solar $\nu_e$'s at 90\% CL.

However, since the $\nu_{e_L}\to \tilde{\nu}_{e_R}$-conversion
within the Sun depends on neutrino energy, sign of $\dot{\Phi}$, etc.,
and due to the absence of the Kamiokande limit for such process below
$E_{\nu}\lsim 8~MeV$ we can not exclude the presence of
right-handed Majorana neutrinos emitted by the Sun, in particular, for
the low-energy beryllium neutrinos (the line $E_{\nu_e}= 0.862~MeV$ ).
One expects to observe these neutrinos in the Borexino
(1998?) and the Hellaz experiments where a
scintillation detector\cite{Borexino} and a helium gas at azote
temperature\cite{Hellaz} should
have a low-energy threshold.

Notice that for the beryllium neutrino line emitted by the Sun the
reaction $\tilde{\nu}_ep\to ne^+$ vanishes because of its high
threshold $E_{\nu th} = 1.8~MeV$, but the
parallel branch--$\tilde{\nu}e$-scattering-- contributes to the
counting of neutrino events starting from a detector threshold.

In anyway, since we assume here the right-handed neutrinos are
producing in the Sun and the first scenario \cite{APS} realizes,
there appear some known objections concerning a large value $\mu
B_{\perp}$ for an effective neutrino spin reversal or non-observation
of the semiannual flux variations in the
current neutrino experiments.

Really, from the astrophysical bound on any Dirac and Majorana neutrino
magnetic moments $\mu_{\nu}\lsim 3\times 10^{-12}\mu_B$\cite{Raffelt}
it follows that necessary neutrino spin reversal
when the dimensionless parameter $\mu_{\nu}B_{\perp}L$ should be of the
order unity, $\mu_{\nu}B_{\perp}L\sim 1$ , is provided by the
sufficiently strong magnetic fields $B\sim 100~kG$ applied along a
half of the width of the convective zone, $L\sim 10^{10}~cm$. Even such
 magnetic fields would not be in contrary with some estimates
in magnetic hydrodynamics\cite{Hughes,Solov'ev}. Meanwhile,
a more weakened and model-independent laboratory constraint
$\mu_{12}\lsim 10^{-10}\mu_B--10^{-11}\mu_B$\cite{Kuld} seems to be
reasonable for a Majorana neutrino for which the SN1987A energy loss
argument\cite{Barbieri1} does not work in contrast to the Dirac case.
This means that magnetic field strength in the Sun could be less than
for the Dirac case, or quite realistic values $B\sim 10^3-10^4~G$ would
be sufficient for effective spin-flavor conversions.

 Non-observation of the semiannual flux variations,
perhaps, is explained either by unsuccessful sign of the angular
velocity $\dot{\Phi}$ for the twisting magnetic field in both solar
hemispheres (in mechanism \cite{APS} under consideration, see above))
or even more simpler than in \cite{APS} without any specific geometry
of a magnetic field: by the presence of random magnetic fields which
always exist in addition to a regular magnetic field and can have the
same strength of the r.m.s. field $\sqrt{<\tilde{B}^2>}\sim B$.  In
contrast to regular field, random magnetic fields do not disappear at
the equatorial plane while $<\tilde B> = 0$ elsewhere and their scale
$L_0$ is much less than the regular field one.

Thus one can pronounce that experimental bounds at present do not
exclude the possible observation  of the low-energy
$\tilde{\nu}_{e_R}$'s in future solar neutrino experiments .

\section{Scattering of partially polarized neutrinos off electrons}

\subsection{Polarization density matrix}

Both right-handed Majorana (active) neutrinos and
right-handed (sterile) Dirac neutrinos can influence the observed
neutrino events but in different ways.

Really, below we find that for partially polarized electron
neutrino, $\mid \xi_z^{(0e)}\mid < 1$, where
\begin{equation}
\rho_{rr'} = \frac{A\delta_{rr'} + \xi_i^{(0e)}(\sigma_i)_{rr'}}{2}
\label{densitymatrix}
\end{equation}
is the spin density matrix of the solar neutrino flux ($r,r'
=\pm 1$) with the initial momentum directed along z-axis, {\bf k}$_1$=
$(0,0,k_1)$, the differential cross-sections of weak interactions,
$(d\sigma/dT)_{weak}$, are the different ones for Dirac and Majorana
neutrinos.

The main point is that the difference of the cross-sections gives the
different electron energy spectrum profiles for two kinds of neutrino.

As
well as in the case of a Dirac neutrino in \cite{Barbieri} the
polarization components of electron Majorana neutrinos in
Eq.~(\ref{densitymatrix}) are\footnote{We use the opposite definition of
the longitudinal polarization to the choice in \cite{Barbieri}, so that
for fully polarized neutrinos we mean $\xi^{(0e)}_z = - 1$ (left-handed
Majorana or Dirac neutrinos).},
\begin{equation}
\begin{array}{ll}
\xi^{(0e)}_z = &\tilde{\nu}_{e_R}^*\tilde{\nu}_{e_R} -
\nu_{e_L}^*\nu_{e_L},\\ \xi_{\perp}^{(0e)} = &2\mid
\nu_{e_L}^*\tilde{\nu}_{e_R}\mid,
\label{components}
\end{array}
\end{equation}
with the essential
distinction that now the right-handed component $\tilde{\nu}_{e_R}$ is
active.

The spin density matrix for the muon neutrinos is given by the same
Eq.(\ref{densitymatrix}) with the change of the index $(0e)$ to $(0\mu )$
and of the subscripts $eL$, $eR$ to $\mu L$, $\mu R$ for the
polarization components (\ref{components})~.

Notice also that, in contrast to \cite{Barbieri}, the normalization of
the electron spin density matrix Eq.~(\ref{densitymatrix}), in general,
is different from unity , \begin{equation} Tr \rho = A =
\tilde{\nu}_{e_R}^*\tilde{\nu}_{e_R} + \nu_{e_L}^*\nu_{e_L} = 1 -
\nu_{\mu_L}^*\nu_{\mu_L} - \tilde{\nu}_{\mu_R}^*\tilde{\nu}_{\mu_R}.
\label{normaliz}
\end{equation}

Using Eq.~(\ref{densitymatrix}) and
summing over the helicities of the partially polarized initial neutrino
flux coming from the Sun to the detector,
$$
\rho =\sum_{r,r'= \pm
`1}u^{r'}_{\nu_e}(k_1)\bar{u}^r_{\nu_e}(k_1)\rho_{rr'},
$$
we obtain in the
ultrarelativistic limit $\omega_1\approx k_1\gg m_{\nu}$ the well-known
density matrix in $4\times 4$-representation (compare
formula (D.54) in the book \cite{Bilenki} for A= 1)
\begin{equation}
\rho \simeq \frac{1}{2}(A + \xi^{(0e)}_z\gamma_5
+\gamma_5\hat{\xi}^{(0e)}_{\perp})\hat{k}_1,
\label{densitymatrix1}
\end{equation}
which allows us to
calculate in the subsection 3.2 the different mean $\nu e$-scattering
cross-sections averaged over solar neutrino spectrum, $<\sigma^{(M)}>$
and $<\sigma^{(D)}>$.

Recall that the index $(0)$ refers to the rest frame of a particle,
$a_{\mu} = (0,\vec{\xi}^{(0)})$, or $\mid \vec{\xi^{(0)}}\mid$ is the
Lorentz-invariant due to the normalization of the spin 4-vector
$a_{\mu}= Tr(\rho\gamma_{\mu}\gamma_5)/2m_e$, $a_{\mu}a^{\mu} = -
\vec{\xi^{(0)}}^2$, in the Mishel-Wightman density matrix $\rho =
(\hat{k}_1 + m_{\nu})(1 + \gamma_5\hat{a})/2$ which transits to
(\ref{densitymatrix1}) for $A = 1$ in the ultrarelativistic limit
$m_{\nu}\to 0$ \cite{Bilenki}.

There are also two additional density matrices with the mixed flavors
in the interference of the weak and electromagnetic amplitudes,

\begin{equation}
\begin{array}{ll}
&\rho^{(\mu \tilde{e})} = \sum_{r, r'}u_{\nu_{eR}}^r(k_1)
\bar{u}_{\nu_{\mu L}}^{r'}(k_1) \rho^{(\mu \tilde{e})}_{rr'}
\simeq \frac{1}{4}\hat{\xi}_{\perp}^{(\mu \tilde{e})}\hat{k}_1(1 + \gamma_5)~,\\

&\rho^{(\tilde{\mu} e)} = \sum_{r, r'}u_{\nu_{\mu R}}^r(k_1)
\bar{u}_{\nu_{e L}}^{r'}(k_1) \rho^{(\tilde{\mu} e)}_{rr'}
\simeq \frac{1}{4}\hat{\xi}_{\perp}^{(\tilde{\mu} e)}\hat{k}_1(1 +
\gamma_5)~, \\

\label{mixture} \end{array} \end{equation}
where the transversal polarization values are given by
\begin{equation}
\begin{array}{ll}
&\mid \vec{\xi}^{(\mu \tilde{e})}_{\perp}\mid  = 2\mid \nu_{\mu L}\mid
\mid \nu_{eR}\mid = 2 \sqrt{P_1(A - P)}\\ &\mid \vec{\xi }^{(e
\tilde{\mu})}_{\perp}\mid = 2\mid \nu_{e L}\mid \mid \nu_{\mu R}\mid =
2\sqrt{P(1 - A - P_1)}~.
 \end{array} \label{mixturepol} \end{equation}

We have determined here the mixed transversal polarizations
(\ref{mixturepol}) as well as the electron neutrino polarization
components  $\xi^{(0e)}_z = A - 2P$,
$\mid \vec{\xi}_{\perp}^{(0e)}\mid  = 2\sqrt{P(A - P)}$ and the muon
neutrino ones $\xi^{(0\mu )}_z = 1 - A - 2P_1$, $\mid
\vec{\xi}_{\perp}^{(0\mu )}\mid  = 2\sqrt{P_1(1 - A - P_1)}$
entering Eq.~(\ref{components}) as the
functions of the survival probability $P = \nu_{e_L}^*\nu_{e_L}$, of
the left-handed muon neutrino probability $P_1 = \nu_{\mu L}^*\nu_{\mu
L}$ and of the normalization coefficient $A$ given by
Eq.~(\ref{normaliz}).  All functions $P, ~P_1$ and $A$ depend on the
neutrino energy $E$, the mixing parameters $\delta s_2$, the magnetic
energy $\mu B_{\perp}$ and the matter density.

These functions are very complicated in the scenario\cite{APS}. For
instance, the merging of the three transitions,
$\nu_{e_L}\leftrightarrow \nu_{\mu_L}$,
$\tilde{\nu}_{e_R}\leftrightarrow \nu_{\mu_L}$,
$\tilde{\nu}_{e_R}\leftrightarrow \nu_{e_L}$, yields the large
$\nu_{e_L}\leftrightarrow \tilde{\nu}_{e_R}$-transition probability
given by Eq. (48) in \cite{APS} ,
$P_{\nu_{e_L}\to \tilde{\nu}_{e_R}} =  A - P$, which after
the averaging over fast oscillations gives
\begin{equation}
P_{\nu_{e_L}\to \tilde{\nu}_{e_R}} = \simeq
\sin^2\frac{4(s_2\delta)^2(\mu B_{\perp})^2}{[(s_2\delta)^2 + (\mu
B_{\perp})^2]^2}.
\label{transition}
\end{equation}
For the beryllium line $E_0 = 0.862~MeV$ the averaging of the
cross-sections with the help of the $\delta$-function, $\sim
\delta (\omega_1 - E_0)$, takes off the energy dependence in the
probabilities like Eq.~\ref{transition}, so that the transition
probabilities and the polarization vector Eq.~\ref{components} depend on the
fundamental vacuum constants, $\Delta m^2$, $s_2$ and on the magnetic
field parameter $\mu B_{\perp}$ which is changing
very slowly comparing with the event counting at the underground
detectors.
Use of $\delta (\omega_1 - E_0)$-function means also that
we have neglected the thermal and Doppler broadening of the beryllium
line \cite{Bahcall}. As result in the case of the beryllium neutrinos we
can parameterize the recoil electron spectra over the values $P=P(\mu
B_{\perp})$ and $A=A(\mu B_{\perp})$ independently on neutrino energy
as well as for the Dirac neutrino conversions $\nu_{e_L}^{(D)}\to
\nu_{e_R}^{(D)}$\cite{Barbieri} which do not depend on neutrino energy
at all.

\subsection{Scattering of Majorana and Dirac neutrinos off electrons}
\vskip 0.5cm

Now using the spin density matrices
Eq.~(\ref{densitymatrix1}),~Eq.~(\ref{mixture}) with the given
polarization components Eq.~(\ref{components}),~Eq.~(\ref{mixturepol})
let us consider what kind of differences of the Majorana and Dirac
neutrino scattering off electrons can arise in underground experiments.

 The key argument here is the difference of the matrix
element for the Majorana neutrino current ,
\begin{eqnarray}
<k_2r_2\mid N(\hat{\bar{\Psi}}^{(M)}\gamma_{\mu}\frac{(1 -
\gamma_5)}{2}\hat{\Psi}^{(M)})\mid k_1,r_1> = &&
-\bar{u}^{r_2}(k_1)\gamma_{\mu}\gamma_5u^{r_1}(k_1) = \nonumber \\
&&=\bar{v}^{r_1}(k_1)\gamma_{\mu}\gamma_5v^{r_2}(k_2),
\label{Majorcurrent}
\end{eqnarray}
from the corresponding one for the Dirac neutrino
($\bar{u}^{r_2}(k_2) (\gamma_{\mu}(1 - \gamma_5)/2)u^{r_1}(k_1)$) or
for the Dirac antineutrino ($\bar{v}^{r_1}(k_1)(\gamma_{\mu}(1 -
\gamma_5)/2)v^{r_2}(k_2)$).

Absence of the vector part in the matrix element Eq.~(\ref{Majorcurrent})
is crucial for the neutrino interactions with particles in the case
of partial polarization,  and it is the direct consequence of the
charge-conjugation property $C\hat{\Psi}^{(M)}C^{-1}=\hat{\Psi}^{(M)}$
for the second-quantized Majorana field\cite{Commins}
\begin{equation}
\hat{\Psi}^{(M)}(x) = \sum_{\vec{p}r}\frac{1}{\sqrt{2E_p}}
\Bigl \{c^r_{\vec{p}}u^r(p)e^{-ipx} + c^{r+}_{\vec{p}}v^r(p)e^{ipx}\Bigr
\}.
\label{field}
\end{equation}
The matrix element for the $\nu e$-scattering $M = M_{weak} + M_{em}$
consists of the two terms for a Majorana neutrino,
\begin{eqnarray}
M_{weak}(\nu_ie^-\to \nu_ie^-) = &&-
2G_F\sqrt{2}\Bigl
[\bar{u}^{r_2}(k_2)\gamma^{\alpha}\gamma_5u^{r_1}(k_1)\Bigr ]\times \nonumber\\
&&\times \Bigl [\bar{u}_e(p_2)(g_{iL}\gamma_{\alpha}\frac{(1 - \gamma_5)}{2} +
 g_R\gamma_{\alpha}\frac{(1 + \gamma_5)}{2})u_e(p_1)\Bigr ],
\label{weak}
\end{eqnarray}
and
\begin{equation}
M_{em} = \frac {ie}{q^2}\Bigl
[\bar{u}^{r_2}(k_2)(\mu^{(M)}_{12} -
id_{12}^{(M)}\gamma_5)\sigma_{\alpha \beta}q^{\beta}u^{r_1}(k_1)\Bigr
]\Bigl [\bar{u}_e(p_2)\gamma^{\alpha}u_e(p_1)\Bigr ].
\label{electromag}
\end{equation}
Here $q_{\beta} = (k_2 -
k_1)_{\beta}$ is the momentum transfer; $g_{iL} = \xi \pm 0.5$ and $g_R = \xi$
are the constants of the standard model with the Weinberg angle
parameter $\xi = \sin^2\theta_W = 0.23$, where upper (lower) sign for $g_{iL}$
corresponds to the subscript $i = e$ for electron neutrino ($i = \mu$ for 
muon neutrinos).

We have meant here that in the presence of a flavor mixed
neutrino flux from the Sun described by the spin density matrices
(\ref{densitymatrix1}),~(\ref{mixture}) the $\nu_ie\to \nu_ie$-
scattering weak amplitude $M_{weak}$ conserving helicity (=
chiralities L or R)
adds the coherent electromagnetic amplitude $M_{em}$
of the spin-flavor transition $\nu_{\mu_{ L,~R}}e\to \nu_{e_{R,~L}}e$ 
in the case of electron neutrinos ($i =e$) or of the inversed process 
$\nu_{e_{ L,~R}}e\to \nu_{\mu_{R,~L}}e$ in the case of muon neutrinos, 
$i = \mu$.

Notice that we do
not distinguish the mass eigen-states in the electromagnetic scattering
$\nu_1e\to \nu_2e$ through the transition dipole moments
$\mu_{12},~d_{12}$ from the flavor states in the
weak amplitude Eq.~(\ref{weak}) because of the inclusion of mixing angles
into $\mu_{12},~d_{12}$ and the ultrarelativistic approximation used in
spinors. Thus , we put $m_1 = m_2 =
m_{\nu_e} = 0$ elsewhere except for $\mu_{12},~d_{12}$ themselves or
even do not bother about the limit $m_{\nu}\to 0$ for a large moment
$\mu\sim 10^{-10}\mu_B$ in some known extended models where $\mu_{ij}$
does not depend on neutrino masses.

It is useful to retain
$\gamma_5$ in CP-violating term in Eq.~(\ref{electromag}) ($\sim d_{12}$)
without the obvious change $\gamma_5u^{r_1} = r_1u^{r_1}$ for the
identical chirality and helicity (in ultrarelativistic approximation
used) because of the mixed polarization Eq.~(\ref{densitymatrix1}), or
due to an arbitrary $r_1$ .

Using Eq.~(\ref{densitymatrix1}), Eq.~(\ref{weak}),
Eq.~(\ref{electromag}) we obtain the differential cross-section for the
Majorana neutrino scattering off electrons,
\begin{equation}
\frac{d\sigma^{(M)}}{dTd\phi} = \Bigl
(\frac{d\sigma^{(M)}}{dTd\phi}\Bigr )_{weak} +
 \Bigl (\frac{d\sigma^{(M)}}{dTd\phi}\Bigr )_{em} +
\Bigl (\frac{d\sigma^{(M)}}{dTd\phi}\Bigr )_{int},
\label{total}
\end{equation}
where the weak amplitude Eq.~(\ref{weak}) leads to the sum of contributions of $\nu_ee$- and $\nu_{\mu}e$ -scattering

\begin{eqnarray}
\Bigl (\frac{d\sigma^{(M)}}{dTd\phi}\Bigr )_{weak} =&&
\frac{G_F^2m_e}{\pi^2}\{
 P\Bigl [ g_{eL}^2 + g_R^2\Bigl (1 - \frac{T}{\omega_1}\Bigr )^2 -
\frac{m_eT}{\omega_1^2}g_{eL}g_R\Bigr ] + \nonumber\\
 && +(A - P)\Bigl [g_R^2 +
\Bigl (1 - \frac{T}{\omega_1}\Bigr )^2g_{eL}^2 -
\frac{m_eT}{\omega_1^2}g_{eL}g_R\Bigr ] + \nonumber\\
 && + P_1\Bigl [ g_{\mu L}^2 + g_R^2\Bigl (1 - \frac{T}{\omega_1}\Bigr )^2 -
\frac{m_eT}{\omega_1^2}g_{\mu L}g_R\Bigr ] + \nonumber\\
 && +(1 - A - P_1)\Bigl [g_R^2 +
\Bigl (1 - \frac{T}{\omega_1}\Bigr )^2g_{\mu L}^2 -
\frac{m_eT}{\omega_1^2}g_{\mu L}g_R\Bigr ] \}~.
\label{weak1}
\end{eqnarray}

and the
electromagnetic amplitude Eq.~(\ref{electromag}) gives
\begin{equation}
\Bigl
 (\frac{d\sigma^{(M)}}{dTd\phi}\Bigr )_{em} = \frac{\alpha^2
}{2m_e^2} \left (\frac{\mu_{12}^2 + d_{12}^2}{\mu_B^2}\right )\left
[\frac{1}{T} - \frac{1}{\omega_1}\right ].
\label{em}
\end{equation}
Both these terms in
Eq.~(\ref{total}) do not depend on the azimuthal angle $\Phi$. However,
the last interference term in Eq.~(\ref{total}),

\begin{eqnarray}
&&\Bigl
( \frac{d\sigma^{(M)}}{dTd\phi}\Bigr )_{int} =
\frac{\alpha G_F} {4\sqrt{2}\pi m_eT}\times \nonumber\\
&& \times \Bigl \{ \Bigl (\frac{\mu_{12}}{\mu_B}\Bigr )
\Bigl [(g_{eL} + g_{\mu L} + 2g_R)(\vec{p_2}\cdot (\vec{\xi}^{(\mu \tilde{e})}_{\perp} +
\vec{\xi}^{(e \tilde{\mu} )}_{\perp} ))\left (2 -
\frac{T}{\omega_1}\right ) + \nonumber \\
&& + (g_{\mu L} - g_{e L})(\vec{p_2}\cdot (\vec{\xi}^{(\mu
\tilde{e})}_{\perp} - \vec{\xi}^{(e \tilde{\mu} )}_{\perp}
))\frac{T}{\omega_1}\Bigr ] - \nonumber\\ && - \Bigl
(\frac{d_{12}}{\mu_B}\Bigr )\Bigl [(g_{eL} + g_{\mu L} + 2g_R)\Bigl (\hat{\vec{k}}_1 \cdot
[\vec{p}_2\times (\vec{\xi}^{(\mu \tilde{e})}_{\perp} + \vec{\xi}^{(e
\tilde{\mu} )}_{\perp}) \Bigr ) \left (2 - \frac{T}{\omega_1}\right ) +
\nonumber\\ && + (g_{\mu L} - g_{e L})\Bigl (\hat{\vec{k}}_1 \cdot [\vec{p}_2\times
(\vec{\xi}^{(\mu \tilde{e})}_{\perp} - \vec{\xi}^{(e \tilde{\mu}
)}_{\perp})] \Bigr )\frac{T}{\omega_1}\Bigr ]\Bigr \}~, \label{int}
\end{eqnarray}

depends on
that angle via the sum in braces, or the cross-section can be rewritten
as the product
\begin{eqnarray}
&&\Bigl ( \frac{d\sigma^{(M)}}{dTd\phi}\Bigr )_{int} =
\frac{\alpha G_F}{4\sqrt{2}\pi m_eT} \mid \vec{p}_2\mid \sin
\theta_{p_2k_1}\left [ \Bigl (\frac{\mu_{12}}{\mu_B}\Bigr )\cos \phi +
\Bigl (\frac{d_{12}}{\mu_B}\Bigr )\sin \phi \right ]\times  \nonumber\\
&&\times \Bigl [(g_{eL} + g_{\mu L} + 2g_R)\Bigl (2 - 
\frac{T}{\omega_1}\Bigr )\Bigl (\mid \vec{\xi}_{\perp}^{(\mu \tilde{e})}  +
\vec{\xi}_{\perp}^{(e\tilde{\mu})}\mid \Bigr ) + \nonumber\\
&& + (g_{\mu L} - g_{eL})\frac{T}{\omega_1}\Bigl (\mid 
\vec{\xi}_{\perp}^{(\mu \tilde{e})} - 
\vec{\xi}_{\perp}^{(e\tilde{\mu})}\mid \Bigr )\Bigr ], \label{sum}
\end{eqnarray}
where $\mid
\vec{p}_2\mid\sin \theta_{p_2k_1} = \sqrt{2m_eT(1 - T/T_{max})}$,~
$T_{max} = 2\omega_1^2/(m_e + 2\omega_1)$ is the maximum kinetic energy
of the recoil electron, and the transversal polarization values
Eq.~(\ref{mixturepol}), $\mid \vec{\xi}^{(\mu \tilde{e})}_{\perp}\mid=
2\sqrt{P_1(A - P)}$,~$\mid \vec{\xi}^{(e\tilde{\mu})}_{\perp}\mid=
2\sqrt{P(1 - A - P_1)}$,  are the external parameters which are given
by a neutrino dynamics within solar interior.

The polarization vectors $\vec{\xi}_{\perp}^{(a)}$ should follow the
solar magnetic field direction at the resonant point along neutrino
trajectory obeying
the extremum conditions $d\xi^{(\mu \tilde{e})}_z/dt \equiv d(A - P -
P_1)/dt = 0$, $d\xi^{(e \tilde{\mu})}_z/dt \equiv d(1 - A - P -
P_1)/dt = 0$ or $d\xi^{(0e)}_z/dt \equiv d(A - 2P)/dt = 0$, and the
Bargman-Mishel-Telegdi' equation for the electroneutral fermion spin
motion in the external magnetic field {\bf B}\cite{BLP}, $$
\frac{d\xi^{(a)}_z}{dt} = 2\mu_{\nu}\Bigl (\vec{\xi}^{(a)}_{\perp}\cdot
[\vec{n}\times \vec{B}]\Bigr )~. $$ Here the unit vector {\bf n} =
{\bf k}$_1/\omega_1$ and the neutrino momentum {\bf k}$_1 =
(0,0,\omega_1)$ are orthogonal to the transversal polarization,
$(n_i\cdot \xi_{\perp i}^{(a)}) = 0$, and $a = (0e),~(0\mu )~,(\mu
\tilde{e})~,(e \tilde{\mu })$ is the flavor index.  Since all
transversal polarization vectors are aligned along the magnetic field
{\bf B} we may change  the modulus of the vector difference and sum
$\mid \vec{\xi}^{(\mu \tilde{e})}_{\perp} \pm
\vec{\xi}^{e\tilde{\mu})}_{\perp} \mid$ in the interference
cross-section (\ref{sum}) to the difference and sum
of the vector values $\mid \vec{\xi}^{(\mu \tilde{e})}_{\perp}\mid  \pm
\mid \vec{\xi}^{e\tilde{\mu})}_{\perp} \mid$ given by
Eq.~(\ref{mixturepol})~.

It is obvious that the interference term (\ref{int}) vanishes for the
separated electron ($A = 1$~, $P_1 = 0)$ and muon ($A = P = 0$)
systems.

Let us recall that, in general, the electron neutrino survival
probability $P = P(\omega_1, \mu B_{\perp}, s_2, \Delta m^2)$, the
muon neutrino parameter $P_1 = P_1(\omega_1, \mu
B_{\perp}, s_2, \Delta m^2)$  and the normalization factor $A =
A(\omega_1, \mu B_{\perp}, s_2, \Delta m^2)$ appearing in
Eq.~(\ref{weak1}) Eq.~(\ref{int}),  are some functions of the incident
neutrino energy $\omega_1$, of the solar magnetic field $B_{\perp}$
multiplied by $\mu$, and of the fundamental vacuum parameters $s_2=
\sin 2\theta$, $\Delta m^2 = m_1^2 - m_2^2$.

Analogously to Eq.~(\ref{em}) which is proportional to the factor
$(\mu_{12}^2 + d_{12}^2)$  in the interference cross
section Eq.~(\ref{int}) both terms that are linear in $\mu_{12}$ and
$d_{12}$ have the same CP-signs while the intrinsic CP-signs of these
dipole moments are opposite.  Actually, in Eq.~(\ref{int}) this
coincidence of CP-signs is due to the different CP-signs of the kinematic
factors before transition moments, or due to the presence of the third
additional 3-vector $\hat{k}_1$ before $d_{12}$. On the other hand, if
CP-holds:  $(CP)\cal{L}$$_{int}(CP)^{-1}$$=\cal{L}$$_{int}$, one term
survives in Eq.~(\ref{em}) and Eq.~(\ref{int}) only: either $\sim
\mu_{12}$ or $\sim d_{12}$.

Notice that for the case $d_{12} = 0$ the cross-section Eq. (12) coincides 
after a change of notations with the recent result \cite{Gaida} obtained 
by I.V. Gaidaenko.

Let us compare the result Eq.~(\ref{total}) with the analogous
Eq.(11) obtained in \cite{Barbieri} for the Dirac neutrino conversions
$\nu_{e_L}\leftrightarrow \nu_{e_R}$ in the Sun with the following
neutrino scattering off electrons in underground detectors. The
electromagnetic term Eq.~(\ref{em}) is the same as in the Dirac
case with the trivial change $\mu_{\nu}^{(D)}\to \sqrt{\mu_{12}^2 +
d_{12}^2}$ for a Majorana neutrino. However, the main term,
Eq.~(\ref{weak1}), and the interference one Eq.~(\ref{int}) have
electron energy spectra which are different from the spectra for the
corresponding terms in \cite{Barbieri}.

Really, repeating calculations of \cite{Barbieri} with use of
Eq.~(\ref{densitymatrix1}) for the separated
$\nu_{e_L},~\nu_{e_R}$-system ($A = 1$) we find \begin{equation} \Bigl
(\frac{d\sigma^{(D)}}{dTd\phi}\Bigr )_{weak} =
P\frac{G_F^2m_e}{\pi^2}\Bigl [g_L^2
+ g_R^2\Bigl (1 - \frac{T}{\omega_1}\Bigr )^2 -
\frac{m_eT}{\omega_1^2}g_Lg_R \Bigr ],
\label{weak2}
\end{equation}
and
\begin{equation}
\Bigl (\frac{d\sigma^{(D)}}{dTd\phi}\Bigr )_{int} = -
(\vec{p_2}\cdot \vec{\xi}^{(0e)}_{\perp}) \frac{\mu_{\nu}^{(D)}}{\mu_B}
\frac{\alpha G_F}{2\sqrt{2}\pi m_eT}\left (g_L +
 g_R\left [1 - \frac{T}{\omega_1}\right ]\right ),
\label{int1}
\end{equation}
where the last interference term is exactly Eq. (11c) in
\cite{Barbieri},
but the weak interaction term Eq.~(\ref{weak2}) is two times bigger than
Eq. (11a) in \cite{Barbieri}.

Our cross-sections $(d\sigma^{(M,D)}/dT)_{weak}$
Eq.~(\ref{weak1}), Eq.~(\ref{weak2}) coincide with the standard results
\cite{Okun} for the scattering of the massless, fully-polarized
neutrinos.  For instance, for the separated
$\nu_{e_L},~\tilde{\nu}_{e_R}$-system (A = 1) this coincidence,
($d\sigma^{(M)}/dT)_{weak} = d\sigma^{(D)}/dT)_{weak}$, is shown in
Fig. 1 by the common line "a" which describes both the cross section of
the $\nu_ee\to \nu_ee$-scattering for the left-handed Majorana
neutrinos with the substitution $\xi^{(0e)}_z = - 1$ (A = P = 1) to
Eq.~(\ref{weak1}) and the cross section Eq.~(\ref{weak2}) for the
fully-polarized Dirac neutrinos with the substitution P =1.

One can also check from Eq.~(\ref{weak1}) that the
right-handed fully polarized Majorana neutrinos, $\xi^{(0e)}_z = 1$, A =
1, P = 0, are described by the standard cross-section for the Dirac
antineutrinos ($g_L\leftrightarrow g_R$ and $P\to 1$ in
Eq.~(\ref{weak2}) which, in contrast to the Majorana
$\tilde{\nu}_{e_R}$, can not be observed in solar neutrino flux (line
"d" in Fig.  1).

Moreover, in the particular case $A = 1$ and after a trivial transition
from CM to the laboratory frame of reference we have found the full
coincidence of our Eq.~(\ref{weak1}) with the old result Eq. (3.9) in
\cite{Shrock}.

If a sharp difference in the slope of lines for the Majorana neutrinos
("b", "c", "d" lines in Fig.1) and for the Dirac ones ("e", "f") was
marked in the Borexino or Hellaz experiment this would be a test whether
the solar neutrinos are the Majorana particles or they are the Dirac
ones.  {\it It does not matter which concrete value of unknown
parameter $\mu B_{\perp}$ influencing the survival probability P occurs
within the Sun}: all corresponding lines given in Fig.1 for the same
probabilities (pairs ("b", "e") and ("c", "f")) have quite different
slopes.

Emphasize that we have simplified the observable averaged cross
sections\\
$<d\sigma/dT>$ using $\delta (\omega_1 - E_0)$-function for
the beryllium neutrinos. All lines are interrupted at $T_{max}\simeq
0.663~keV$ (see also Fig.2,3) and authors\cite{Borexino} are planning
to distinguish the "signature edge" of the recoil spectrum from the
monoenergetic Be neutrino flux.

This difference appears also for a non-separated
system (merging of resonances in the mechanism\cite{APS}) shown in Fig.
2 for the particular equipartition case $P = \nu_{e_L}^*\nu_{e_L} =
\tilde{\nu}_{e_R}^*\tilde{\nu}_{e_R} =\nu_{\mu_L}^*\nu_{\mu_L} =
\tilde{\nu}_{\mu_R}^*\tilde{\nu}_{\mu_R} = 0.25$ when the
$\nu_{\mu}e$-scattering is taken into account.

The similar difference between Majorana and Dirac neutrino cases
can be found from the total cross section behavior shown in Fig.3
for the separated electron neutrino system (A=1).  These cross sections
$\sigma (W_e)$ contribute to the total number of neutrino events which
can be observed in some different energy bins separated by the
different thresholds $W_e$.  For a low energy region all lines
depending on the different parameters P which correspond to the
Majorana neutrino cross sections become parallel to each other (lines
"b", "c" in Fig.  3) but the analogous Dirac neutrino cross-sections
(lines "d", "e" with the same P as in the Majorana case) change their
slope. Thus, if the magnetic field value $B_{\perp}$ in the Sun is
somehow changing and, therefore, influences the total cross section
$\sigma^{(D,M)}_{\nu e}(W_e)$ (i.e.  leads to the transition $"b"\to
"c"$, or $"d"\to "e"$) we can distinguish Majorana and Dirac
neutrinos due to different slopes of the event spectra over $W_e$,
$N_{\nu}(W_e)\sim <\sigma^{(D,M)}_{\nu e}(W_e)>$.

The resonant spin precession can appear for another changing magnetic
 field configuration, for instance, for the {\it linear -polarized
 Alfven wave} \cite{Semikoz} when a magnetic field direction
(transversal to the incident neutrino momentum) could be fixed. However,
in contrast to \cite{Akhmedov,APS}, this mechanism is still not
elaborated for $\nu_{e_L}\to \tilde{\nu}_{\mu_R}$ and $\nu_{e_L}\to
\tilde{\nu}_{e_R}$-conversions.

\section{Discussions and conclusions}
\vskip 1truecm
The partially polarized electron neutrino flux provided by $\nu_{e_L}\to
\tilde{\nu}_{e_R}$ conversions within the Sun can give an opportunity to
distinguish Majorana and Dirac neutrinos in the Borexino and in
the Hellaz experiments due to quite different profiles of recoil
electron spectra(Fig.1-3). Notice that measurements of $T$ and
$\theta_{p_2k_1}$ would allow authors of the project\cite{Hellaz} to
determine the neutrino energy $\omega_1$.

The inclusion of other neutrino fluxes with the different $\Phi_i
(\omega_1)$ is necessary for comparison  with real electron
spectra although we expect the same qualitative difference of the
spectra for the Majorana and the Dirac cases.

Really, it is obvious that $d\sigma^{(M)}/dT$
exceeds $d\sigma^{(D)}/dT$ in a wide energy region  due to
additional interaction of right-handed {\it active} Majorana neutrinos
with electrons in contrast to the {\it sterile} $\nu^{(D)}_R$.

Probably, in the Superkamiokande or in the SNO detectors this
difference can be marked too.  In spite of unknown dependence
of the survival probability $P(\omega_1, \mu B_{\perp})$ on the
incident neutrino energy $\omega_1$ both the differential
$<d\sigma/dT>$ and the total $<\sigma (W_e)>$ cross-sections averaged
over the boron neutrino spectrum $\lambda (\omega_1)$\cite{Bahcall}
should have quite different slopes depending on $T$ or $W_e$~. One
can easily check this statement substituting a fixed probe value of
$0\leq P< 1$ and integrating Eq.~(\ref{weak1}),
Eq.~(\ref{weak2}) over the spectrum $\lambda (\omega_1)$\cite{Bahcall}.
The absolute value of the averages $<d\sigma/dT>$, $<\sigma (W_e)>$ 
changes when $P$ varies that corresponds to some
different levels of the electron neutrino deficit. Meanwhile both sets
of lines (for Majorana or Dirac neutrinos) conserve different slopes
for the two kinds of particles.

Another model independent way to distinguish Majorana and Dirac
neutrinos was proposed in \cite{Bilenky}. Authors \cite{Bilenky}
showed that NC events at the SNO detector would not oscillate over
time for $\nu^{(M)}$ and would oscillate in the case of $\nu^{(D)}$.
The method is based on any active-active transitions $\nu_a\to
\tilde{\nu}_b$ in the solar magnetic field for Majorana neutrinos and
on active-sterile conversions in the Dirac case. A changing magnetic
field $B(t)\neq 0$ influences neutrino flux variations so that one
predicts $dN^{NC}/dt = 0$ for $\nu^{(M)}$ and $dN^{NC}/dt \neq 0$ for
$\nu^{(D)}$.

While in the method \cite{Bilenky} one expects $dN^{NC}/dt = 0$ for
both kinds of neutrinos if $B\simeq const$, in our method in
a constant magnetic field the profiles $<d\sigma^{(M)}/dT>$ and
$<d\sigma^{(D)}/dT>$ remain different.

In conclusion, without any knowledge of solar magnetic field
 direction and even for a magnetic moment obeying the astrophysical
constraint $\mu\lsim 3\times 10^{-12}\mu_B$ \cite{Raffelt} one
can distinguish Majorana and Dirac neutrino measuring the recoil
electron spectra in the low energy $\nu_ee$-scattering in underground
detectors and comparing slopes of their profiles.

Last remark concerns a possibility to observe some variations of
neutrino flux in detectors with large statistics of events.
Let us assume that the unknown parameter $\mu B_{\perp}$
can change somehow in the Sun influencing $P(\mu B_{\perp})$. In this
case the spectra $<d\sigma^{(M)}/dT>$ will be displaced parallel to the
initial one without change of a slope in the contrary to the spectra
$<d\sigma^{(D)}/dT>$ that should change its slope. This behaviour of
spectra would be a strong argument in favour of the mechanism\cite{APS}
as solution of the solar neutrino problem.
\vskip 0.5cm
\noindent
{\bf Acknowledgment}\\
I acknowledge George Zatsepin for useful discussion and for pointing
out Refs. \cite{Borexino}, \cite{Hellaz} and correspondence with Robert
Shrock  and Samoil Bilenky regarding the Refs. \cite{Shrock1}, \cite{Shrock}
and the Ref. \cite{Bilenky}. 
I am grateful to I.V. Gaidaenko who was first obtaining correct
cross-section of the $\nu e$-scattering of partially polarized 
Majorana neutrinos and indebted to him  for sending of his preprint
\cite{Gaida} that allowed me to improve a wrong previous version
of Eq. (12). 
This work was supported in part from RFFR under grants N. 95-02-03724 
and N. 97-02-16501.


\newpage
\vskip 3cm
\begin{center}
{\bf Figure Captions}
\end{center}
\vskip 1cm
{\bf Fig. 1.}
The differential spectra Eqs.~(\ref{weak1}) and (\ref{weak2}) for the
beryllium neutrino in the case of the separated
$\nu_{e_L},~\tilde{\nu}_{e_R}$--system (the normalization parameter A in
Eq.~(\ref{normaliz}) equals to unity, $A= 1$):

The lines "b", "c", "d" correspond to an incident Majorana
neutrino with the survival probability $P = 0.5~, 1/3,~0$
correspondingly, and the lines "e", "f" describe cross sections for a
Dirac neutrino in the cases $P = 0.5,~ 1/3$. The line "a" is the
common one for Majorana and Dirac fully-polarized left-handed neutrinos
with $P =1$.
\vskip 1cm
{\bf Fig. 2.}
The same spectra in the case of
the equipartition $P =\nu_{e_L}^*\nu_{e_L} =
\tilde{\nu}_{e_R}^*\tilde{\nu}_{e_R} = \nu_{\mu_L}^*\nu_{\mu_L} =
\tilde{\nu}_{\mu_R}^*\tilde{\nu}_{\mu_R} = 0.25$, the parameter $A =
0.5$:

The line "a" describes Majorana neutrino, the line "b"-- Dirac
neutrino.
\vskip 1cm
{\bf Fig. 3.}
The total cross sections for the beryllium neutrino
$<\sigma^{(M,D)}(W_e)>$ in dependence on the threshold energy $W_e$.
For a liquid scintillator it assumes a low threshold $W_e$ of
order $\sim$ a few hundreds keV.
 The line "a" corresponds to the fully polarized left-handed neutrinos
($P = 1$) with $<\sigma^{(M)}(W_e)> = <\sigma^{(D)}(W_e)>$ (SSM
prediction without $\nu_e(Be)$- deficit); lines "b", "c" are plotted
for Majorana neutrino with the survival probabilities $P = 0.5$ and $P
= 1/3$ correspondingly; lines "d", "e" are plotted for Dirac
neutrino with the same $P = 0.5$ and $P = 1/3$.

\newpage





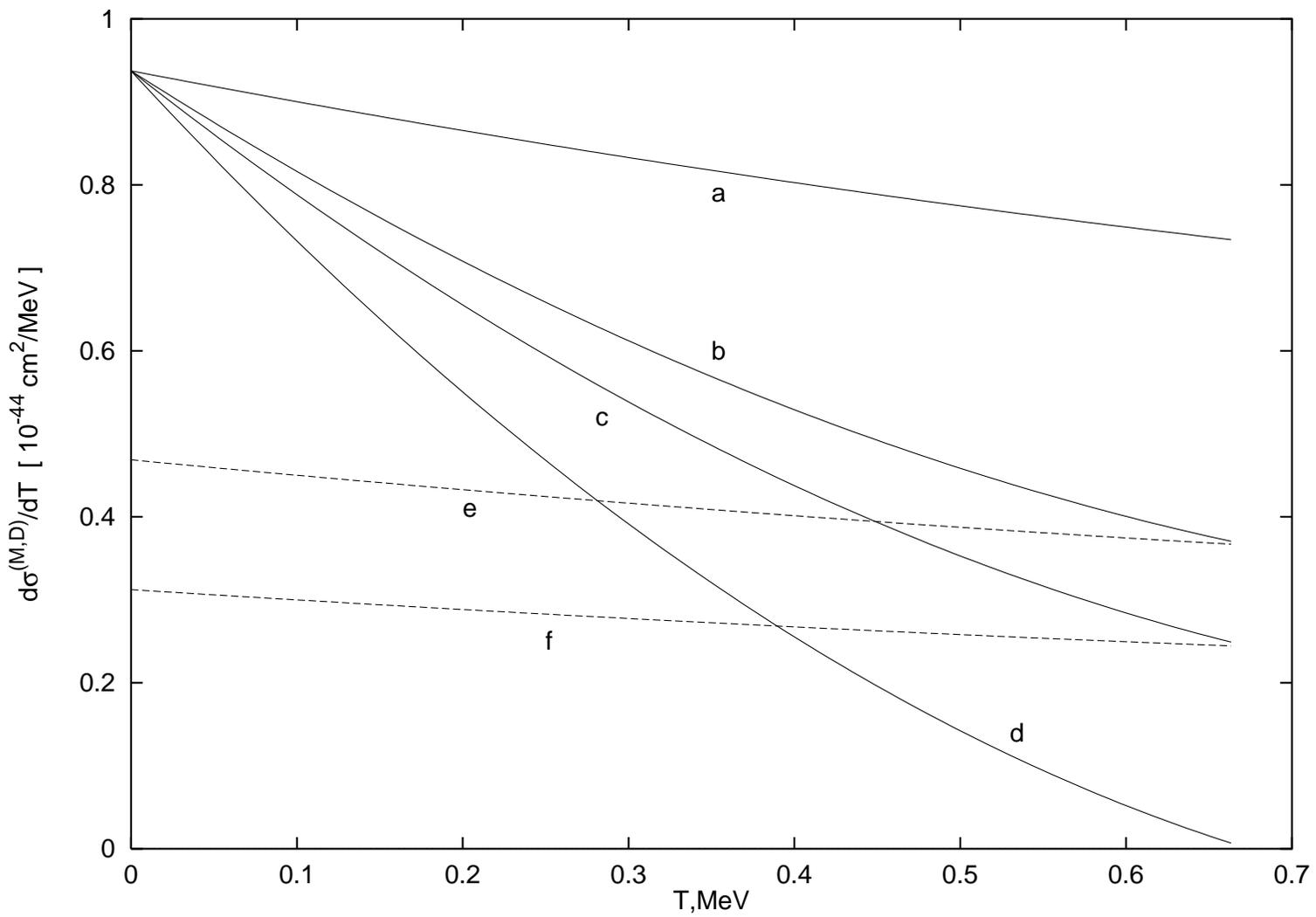
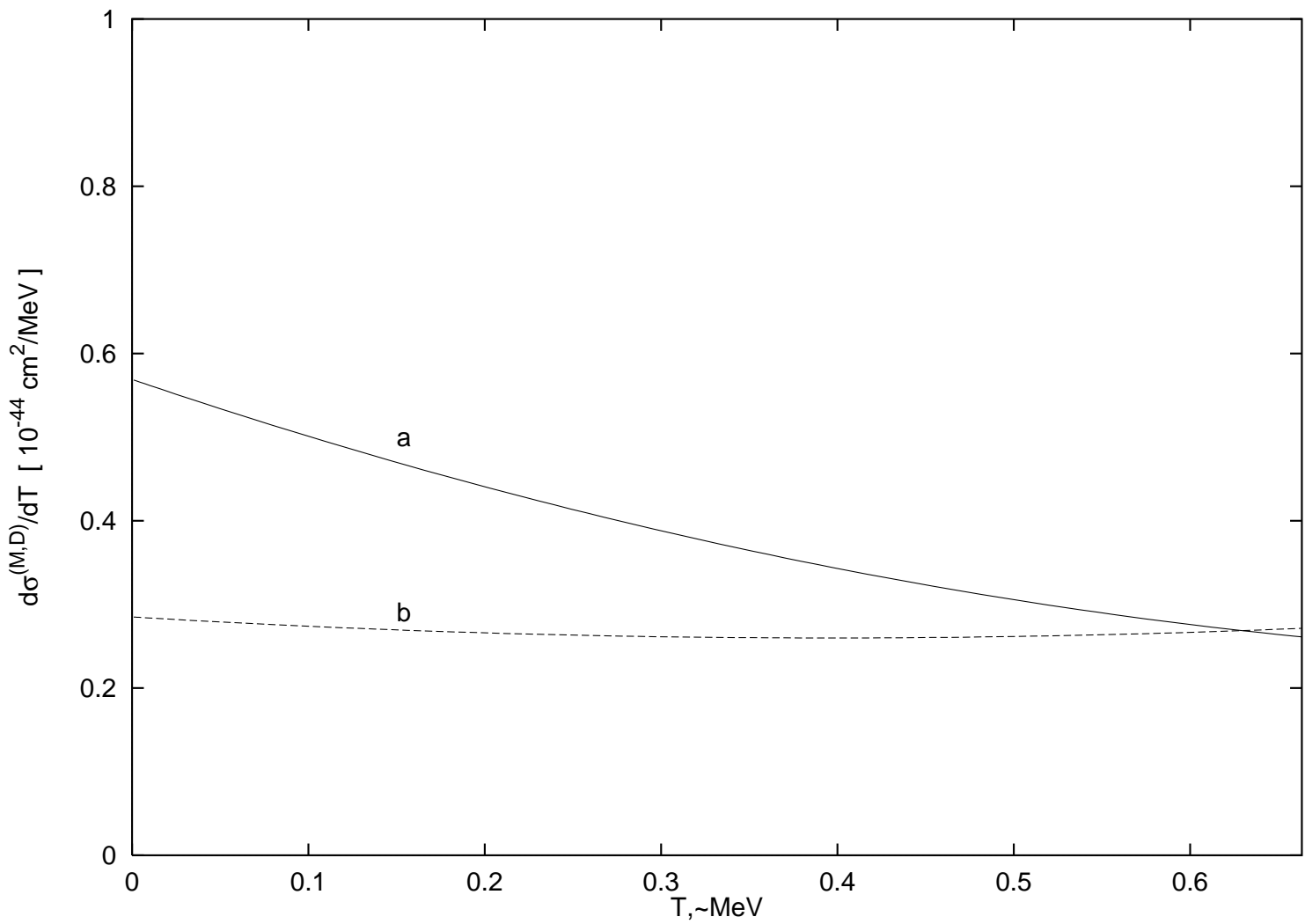
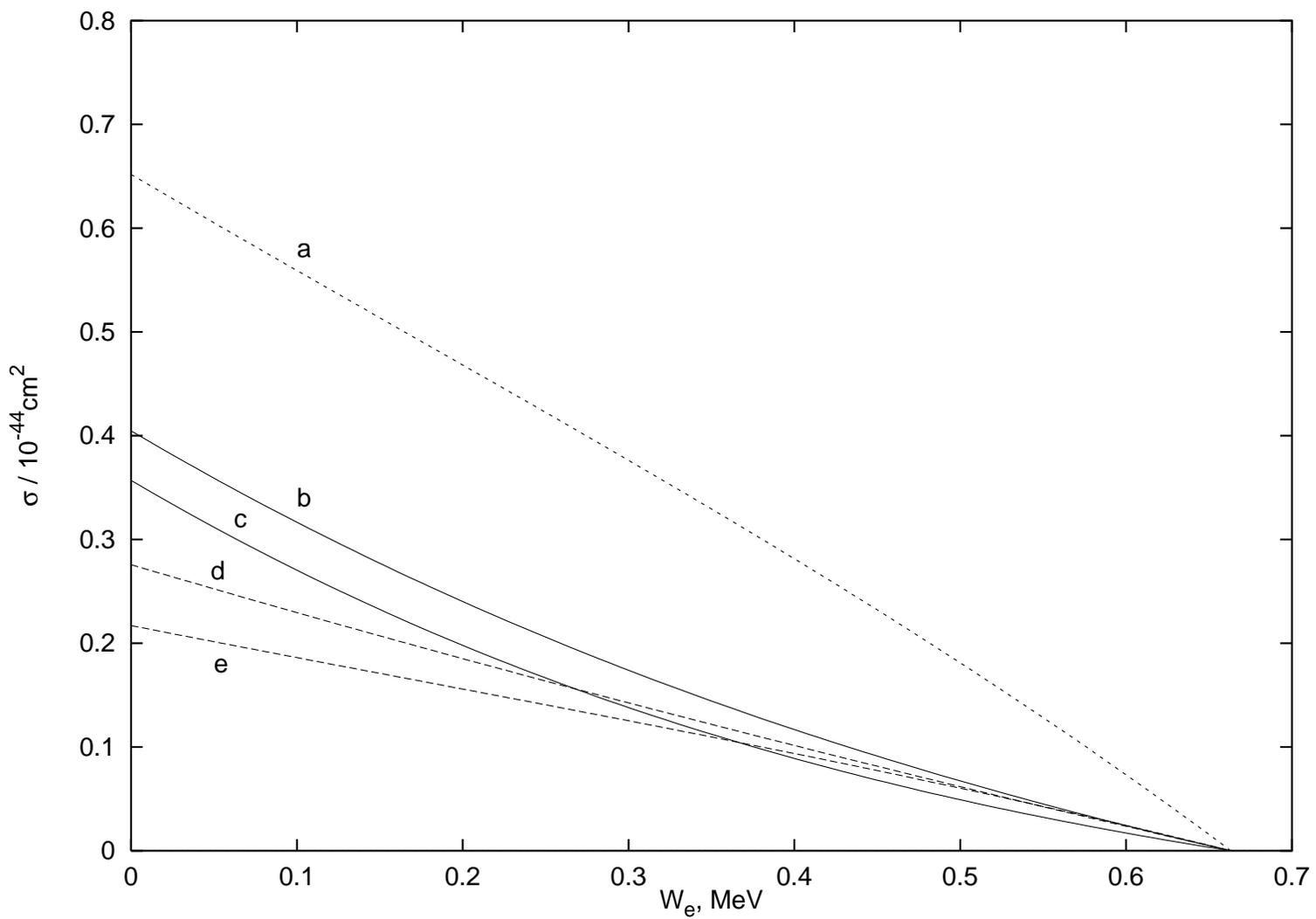
\end{document}